\documentclass[%
 reprint,
 amsmath,amssymb,
 aps,
pra,
]{revtex4-2}

\usepackage{algorithm}
\usepackage{algpseudocode}
\usepackage{float}
\usepackage{xcolor}
\usepackage{etoolbox} 
\usepackage{graphicx}
\usepackage{amsmath}
\usepackage{amssymb}
\usepackage{bm}
\usepackage{hyperref}

\algnewcommand{\algorithmicdoinparallel}{\textbf{do in parallel}}
\algnewcommand{\algorithmicenddoinparallel}{\textbf{end parallel}}
\algdef{SE}[PARALLEL]{Parallel}{EndParallel}[1]
  {\algorithmicdoinparallel\ #1}
  {\algorithmicenddoinparallel}

\begin{document}

\title{GenML: A Python library to generate the Mittag-Leffler correlated noise}

\author{Xiang Qu}
\author{Hui Zhao}
\author{Wenjie Cai}
\author{Gongyi Wang}
\author{Zihan Huang}
 \email{huangzih@hnu.edu.cn}
\affiliation{School of Physics and Electronics, Hunan University, Changsha 410082, China}


\begin{abstract}

Mittag-Leffler correlated noise (M-L noise) plays a crucial role in the dynamics of complex systems, yet the scientific community has lacked tools for its direct generation. Addressing this gap, our work introduces GenML, a Python library specifically designed for generating M-L noise. We detail the architecture and functionalities of GenML and its underlying algorithmic approach, which enables the precise simulation of M-L noise. The effectiveness of GenML is validated through quantitative analyses of autocorrelation functions and diffusion behaviors, showcasing its capability to accurately replicate theoretical noise properties. Our contribution with GenML enables the effective application of M-L noise data in numerical simulation and data-driven methods for describing complex systems, moving beyond mere theoretical modeling.

\end{abstract}

\maketitle

\section{Motivation and significance}

As a ubiquitous stochastic process in complex systems, noise is integral to the evolution of such systems. The intrinsic physical properties of noise are essential for understanding the mechanisms underlying system behaviors \cite{Franosch,Jhawar,Plenio,Ghanta}. The basic theoretical description of noise in various studies is the white noise, which exhibits a time-independent feature and a uniform power spectrum \cite{BERNIDO}. However, unexpected behavior patterns in complex systems are often induced by non-white noise with temporal correlations \cite{Chun,Zhou,Kamenev,Kazakevicius}. Numerous studies have supported this perspective, among which are representative examples: in ultracold and collisionless atomic gases, non-white noise modulates the light scattering growth rate based on its correlation time \cite{Zhou}; in the survival dynamics of biological populations, non-white environmental noise affects the dependency of population extinction times \cite{Kamenev}; and in thermophoretic effects, non-white noise directs the motion of Brownian particles in fluids along the temperature gradient \cite{Kazakevicius}. These findings highlight the pivotal role of non-white noise in revealing the internal workings of complex systems, and emphasize the necessity of considering noise effects when elucidating the evolution dynamics.

In particular, Mittag-Leffler correlated noise (referred to as M-L noise in this paper) is distinguished for its capability to emulate a broad spectrum of correlation behaviors through parameter adjustments \cite{Vinales, Vinales2014}, including typical exponential and power-law correlations. This flexibility makes M-L noise a versatile tool for modeling a wide range of phenomena in complex systems \cite{Vinales,Vinales2014,Fa,Laas,Cairano,Umamaheswaria}. For example,  Laas et al. utilize M-L noise to offer a theoretical exploration of the resonance behaviors observed in Brownian particles within oscillatory viscoelastic shear flows \cite{Laas}. Cairano et al. underscore M-L noise's role in unraveling the crossover from subdiffusive to Brownian motion and the relaxation dynamics within membrane proteins \cite{Cairano}. Umamaheswari et al. demonstrate the application of M-L noise in financial mathematics by leveraging it to explain the existence and stability of solutions in nonlinear stochastic fractional dynamical systems \cite{Umamaheswaria}. That is, the importance of M-L noise in dissecting complex system dynamics is significant, necessitating advancements in its generation for deeper insights.

However, despite significant efforts by researchers to develop algorithms for simulating various types of non-white noise \cite{Milotti,Kasdin,Bykhovsky}, direct algorithms and software for generating M-L noise remain notably absent. Such an absence significantly impedes its application in crucial simulation methods such as Langevin dynamics \cite{Stella,Burov} and molecular dynamics \cite{Hollingsworth}, as well as in data-driven approaches like machine learning \cite{Munoz-Gil,Bo,Qu,Li,Feng}. This limitation confines the utility of M-L noise primarily to theoretical modeling, thereby restricting in-depth exploration of its effects on complex systems. Consequently, there is a great demand for the development of tools capable of accurately simulating M-L noise, which could unlock new insights into the dynamic behaviors of complex systems across various scientific fields.

To address this critical issue, we introduce GenML \cite{genml}, a Python library designed to effectively generate M-L noise in this paper. This software marks a significant advancement, enabling researchers to directly simulate M-L noise and apply it across a wide range of fields. GenML not only fills the existing gap in the available simulation tools for M-L noise but also paves the way for new research opportunities in understanding and modeling the dynamics of complex systems.

\section{Software description}

Before delving into the details of the GenML software, it is essential to introduce the definition of M-L noise. We denote the noise sequence of length $T$ as $\{ \xi(t)\}$ where $t=0,1,2,\ldots,T-1$. The autocorrelation function $C(t)=\langle \xi(0) \cdot \xi(t) \rangle$ of M-L noise sequence satisfies \cite{Vinales}:
\begin{equation}
C(t) = \frac{C}{\tau^{\lambda}} E_{\lambda}\left[-\left({t/}{\tau}\right)^{\lambda}\right],
\label{eq:eq1}
\end{equation}
where
\begin{equation}
\label{eq:eq2}
E_{\lambda}(z) = \sum_{k=0}^{\infty} \frac{z^k}{\Gamma(\lambda k + 1)}.
\end{equation}
In this context, $\langle \cdot \rangle$ denotes the average over independent instances of noise sequences, \( \ E_{\lambda}(z)\) is the Mittag-Leffler function,  \( \tau \) signifies the characteristic memory time, \( C \) denotes the amplitude coefficient, \( \lambda \) is an exponent with \(
0 < \lambda < 2 \), and \( \Gamma(\cdot) \) refers to the gamma function. With the understanding of M-L noise established, we now turn our attention to the software architecture and functionalities of GenML.

\begin{figure}[t]
\centering
\includegraphics[width=7.6cm]{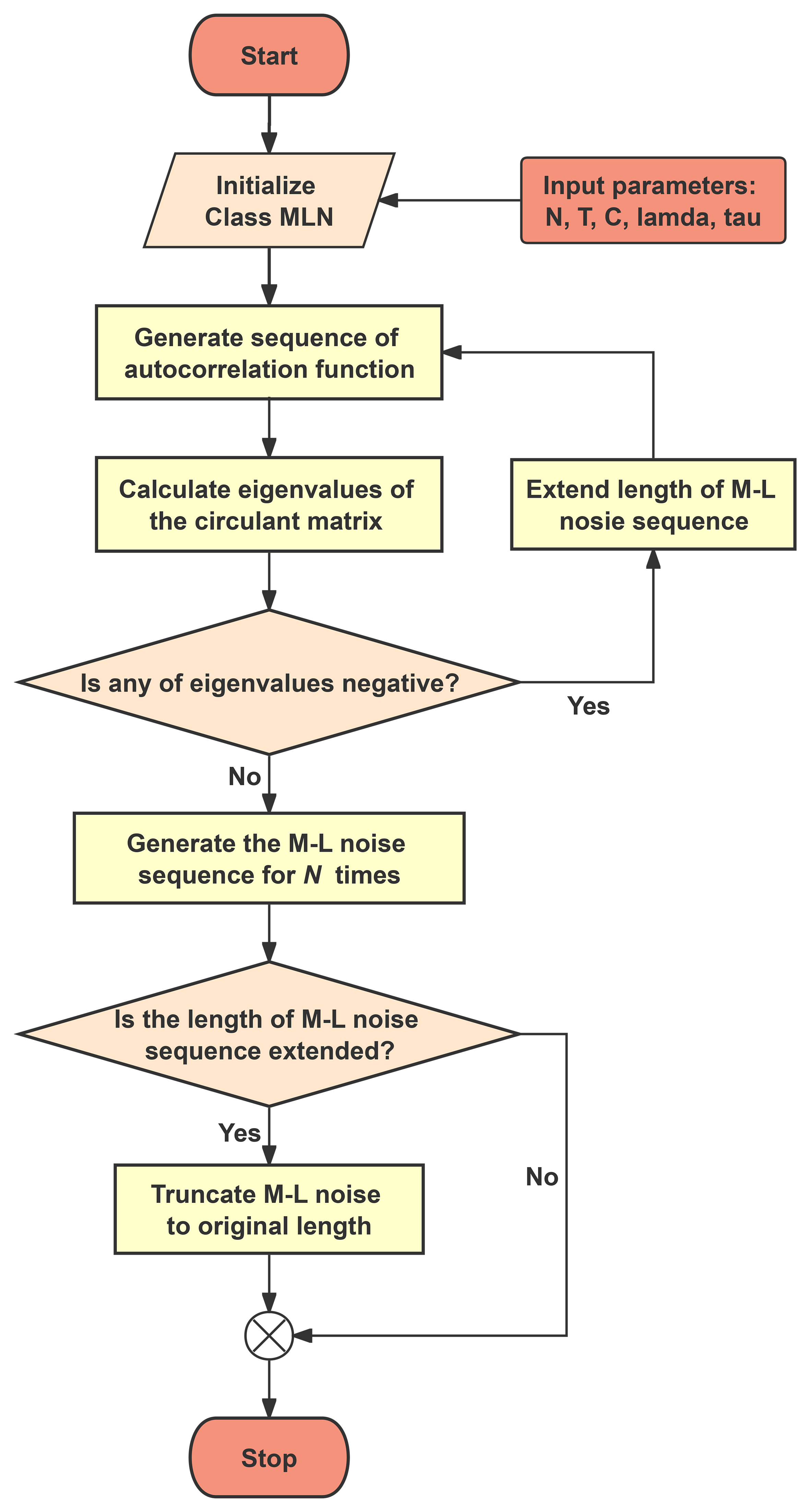}
\caption{Comprehensive workflow of the GenML software for M-L noise generation.}
\label{fig:fig1}
\end{figure}

\subsection{Software architecture}

A comprehensive workflow of GenML is presented in Fig. \ref{fig:fig1}, which is structured into three key steps:

\begin{enumerate}
    \item Determining the optimal sequence length $T_{\rm opt}$ based on the input parameters to ensure the sequence generation algorithm works correctly.
    \item Generating M-L noise sequences of length $T_{\rm opt}$.
    \item Extracting M-L noise sequences of the target length $T$ from the generated sequences.
\end{enumerate}

\begin{figure}[t]
\begin{algorithm}[H]
\caption{Selection of optimal sequence length}
\label{alg:alg1}
\begin{algorithmic}[1]

\State \textbf{Input:}  original length $T$,  autocorrelation function $C(t)$
\State \textbf{Output:} optimal length $T_{\rm opt}$, eigenvalues of corresponding circulant matrix $\{A_k\}$

\State \( \{ c_k \} \leftarrow [c_0, c_1, c_2, \cdots, c_{T-1}, c_{T}, c_{T-1}, \cdots, c_2, c_1] \)
\State \( \{A_k\} \leftarrow \text{FFT}(\{ c_k \}).\text{real} \)
\If{\( \min(\{A_k\}) < 0 \)}
    \For{\( l \) in \( ls \) where \( l > T \)}
        \State \( T \leftarrow l \)
        \State \( \{ c_k \} \leftarrow [c_0, c_1, c_2, \cdots, c_{T-1}, c_{T}, c_{T-1}, \cdots, c_2, c_1] \)
        \State \( \{A_k\} \leftarrow \text{FFT}(\{ c_k \}).\text{real} \)
        \If{\( \min(\{A_k\}) \geq 0 \)}
            \State \( \text{break} \)
        \EndIf
    \EndFor
\EndIf
\State \(T_{\rm opt} \leftarrow T \)
\State \Return $T_{\rm opt}$, $\{A_k\}$

\end{algorithmic}
\end{algorithm}
\end{figure}

Next, we give detailed descriptions of this workflow that empower GenML to accurately generate M-L noise sequences. At first, a sequence of autocorrelation function values is calculated using a numerical evaluation method \cite{Garrappa,mittag-leffler}. For convenience, we denote the value of $C(t)$ in Eq. (\ref{eq:eq1}) as $c_t$, then the sequence $\{c_k\}$ of length $2T$ can be written as:
\begin{equation}
[c_0, c_1, c_2, \cdots, c_{T-1}, c_{T}, c_{T-1}, \cdots, c_2, c_1],
\end{equation}
where $T$ is the noise length. Fast Fourier transform (FFT) is then applied on this sequence. The real part of the FFT results corresponds to the eigenvalues $\{A_k\}$ of the circulant matrix associated with $\{c_k\}$ \cite{Sergej}. These eigenvalues must be non-negative for the subsequent noise sequence generation algorithm. If negative eigenvalues are encountered, the sequence length $T$ should be increased to ensure that the new circulant matrix's eigenvalues are non-negative. However, increasing $T$ also increases the computational resources required, making the identification of an optimal sequence length $T_{\rm opt}$ a critical first step in the workflow. In GenML, a sorted waiting list $ls$ of potential sequence lengths is provided. The software selects the smallest length as $T_{\rm opt} $ from this list that satisfies the requirement for non-negative eigenvalues. Pseudocode of this selection is outlined in Algorithm \ref{alg:alg1}. After determining $T_{\rm opt}$ and $\{A_k\}$, GenML utilizes complex random numbers and the inverse fast Fourier transform (iFFT), referencing the Davies-Harte algorithm \cite{DAVIES,CRAIGMILE}, to generate the M-L noise sequences of length $T_{\rm opt}$. The sequences are then truncated to the original specified length $T$, producing the M-L noise sequences of desired length. Algorithm \ref{alg:alg2} provides a detailed description of this process. Here, different sequences are generated through parallel acceleration. which is achieved using the \texttt{numpy}'s inherent capabilities for parallel operations.

\begin{figure}[t]
\begin{algorithm}[H]
\caption{Generation of M-L noise sequences}
\label{alg:alg2}
\begin{algorithmic}[1]
\State \textbf{Input:} original length $T$, optimal length $T_{\rm opt}$, eigenvalues $\{A_k\}$, number of sequences $N$
\State \textbf{Output:} $N$ M-L noise sequences $\{\xi(t)\}_m$ of length $T$
\Parallel{
\For {\( k = 0, 1, 2, \cdots, 2T_{\rm opt} - 1 \)}
    \State Generate $ Z_k$ {\rm from} ${\rm NormalDistribution}(0,1)$
\EndFor
\For{\( k = 0, 1, 2, \cdots, 2T_{\rm opt} - 1 \)}
    \State Compute \( Y_k \) with \( Z_k \) and \( A_k \) as
    \State \( Y_k \leftarrow \) \( \begin{aligned}
        \begin{cases}\sqrt{2 T_{\rm opt} A_k} Z_0 & k=0 \\
        \sqrt{T_{\rm opt} A_k}\left(Z_{2 k-1}+i Z_{2 k}\right) & 0<k<T_{\rm opt} \\
        \sqrt{2 T_{\rm opt} A_{T_{\rm opt}}} Z_{2 T_{\rm opt}-1} & k=T_{\rm opt} \\
        Y_{2 T_{\rm opt}-k}^* & T_{\rm opt}<k<2 T_{\rm opt}\end{cases}
        \end{aligned} \)
    \State where the asterisk denotes a complex conjugate
\EndFor
\State \( \{\xi(t)\}_m \leftarrow \text{iFFT}(\{Y_k\}) \)
\If {\( T < T_{\rm opt}\)}
    \State truncate $\{\xi(t)\}_m$ to length $T$
\EndIf
\EndParallel}
\State \Return $\{\xi(t)\}_m$ ($m=0,1,2,\cdots,N-1$)
\end{algorithmic}
\end{algorithm}
\end{figure}

\subsection{Software functionalities}

GenML can be conveniently installed using the Python pip command:
\begin{verbatim}
    pip install genml
\end{verbatim}
This software offers an application programming interface (API) to generate M-L noise sequences with different input parameters. The usage of this API is as follows:
\begin{verbatim}
    import genml
    xi = genml.mln(N, T, C, lamda, tau, seed)
\end{verbatim}
where:
\begin{itemize}
  \item \verb|N| represents the number $N$ of M-L noise sequences to be generated.
  \item \verb|T| denotes the length $T$ of noise sequence.
  \item \verb|C| is the amplitude coefficient $C$ of M-L noise.
  \item \verb|lamda| stands for the exponent $\lambda$ of M-L noise, which should be within the range (0,2).
  \item \verb|tau| signifies the characteristic memory time $\tau$, which should be within the range (0,10000].
  \item  \verb|seed| is an integer that seeds the software to ensure reproducibility of the noise sequences, which defaults to ``None". When \verb|seed| is ``None", each call to the \verb|genml.mln| API will generate a different noise sequence. Conversely, setting \verb|seed| to a specific value will ensure the generated noise sequences remain consistent.
\end{itemize}
Upon calling this API, it returns an \verb|ndarray| of shape $(N, T)$, which consists of $N$ M-L noise sequences with specified input parameters.

Another API \verb|genml.acf| is provided in GenML to calculate autocorrelation function values of generated M-L noise sequences. Corresponding usage is given as:
\begin{verbatim}
    acfv = genml.acf(xi, tmax, dt, nc)
\end{verbatim}
where:
\begin{itemize}
    \item \verb|xi| refers to the M-L noise sequences generated by \verb|genml.mln|.
    \item \verb|tmax| denotes the right boundary for calculating the autocorrelation function, which should be less than the length of noise sequence.
    \item \verb|dt| represents the sampling interval, which should be within the range [1, \verb|tmax|).
    \item \verb|nc| is number of CPU cores allocated for parallel computing \cite{McKerns}, with a default setting of 1.
\end{itemize}
This API computes the numerical values of the autocorrelation function based on the definition $C(t)=\langle \xi(0) \cdot \xi(t) \rangle$, storing the results in an \verb|ndarray| list.

\begin{figure}[b]
\centering
\includegraphics[width=8.2cm]{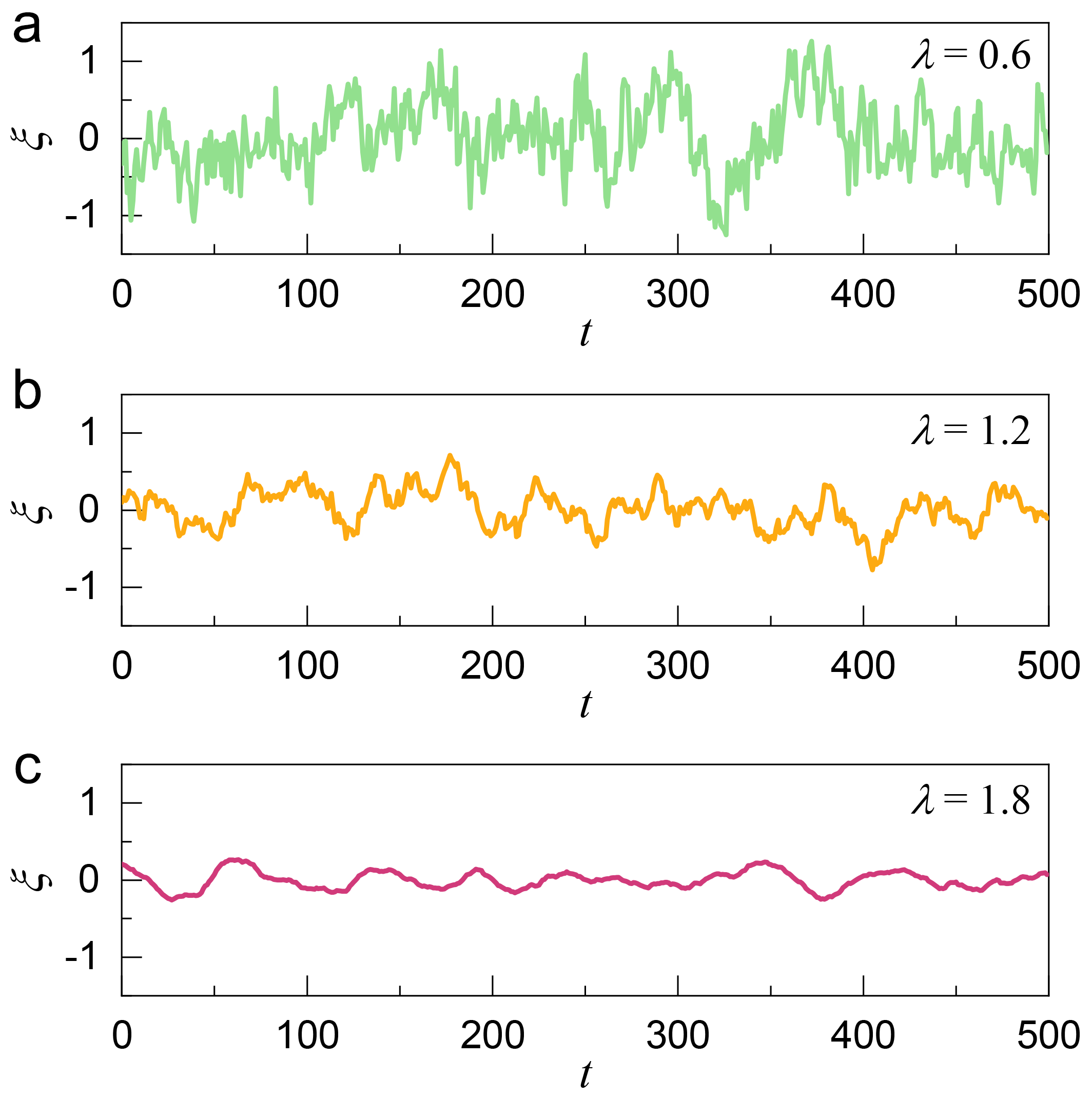}
\caption{Representative examples of M-L noise generated by GenML with $N = 1$, $T = 500$, $C=1$, $\tau = 10$, where $\lambda$ = 0.6 (a), 1.2 (b), and 1.8 (c).}
\label{fig:fig2}
\end{figure}

\begin{figure*}[t]
\centering
\includegraphics[width=16.0cm]{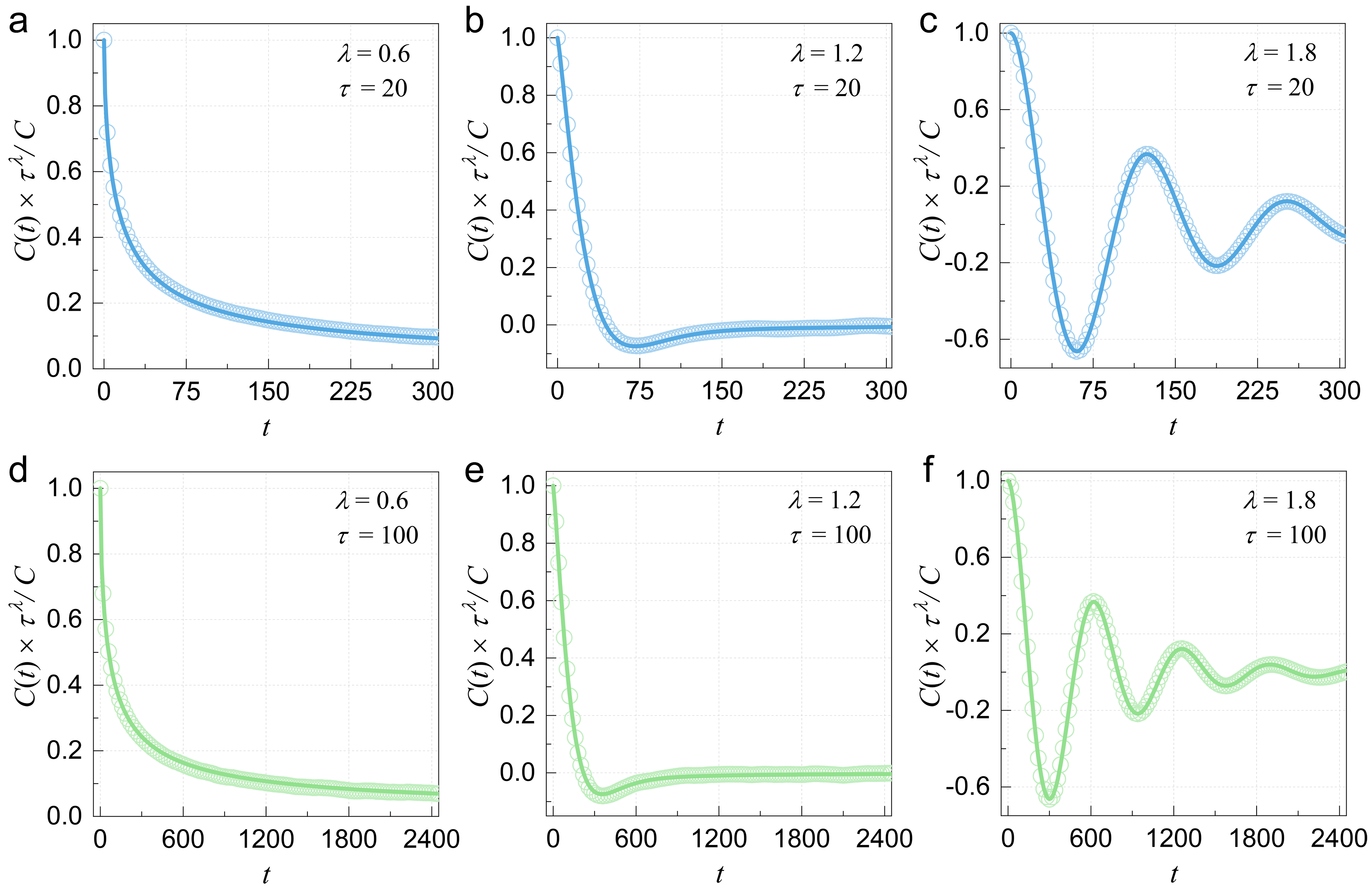}
\caption{Illustrative comparisons of calculated (dots) and theoretical (lines) normalized autocorrelation functions of M-L noise across various parameter sets: (a) $\lambda=0.6$, $\tau=20$; (b) $\lambda=1.2$, $\tau=20$; (c) $\lambda=1.8$, $\tau=20$; (d) $\lambda=0.6$, $\tau=100$; (e) $\lambda=1.2$, $\tau=100$; (f) $\lambda=1.8$, $\tau=100$. The calculated autocorrelation function of each parameter set is based on an average of 1000 independent noise sequences.}

\label{fig:fig3}
\end{figure*}

In addition, to facilitate the comparison between calculated and theoretical values of the autocorrelation function, we also offer the API \verb|genml.acft| for computing the theoretical values of the M-L noise autocorrelation function. Its usage is as follows:
\begin{verbatim}
    acftv = genml.acft(tmax, dt, C, lamda, tau)
\end{verbatim}
Here, the parameters \verb|tmax|, \verb|dt|, \verb|C|, \verb|lamda|, and \verb|tau| retain the same definitions as those in \verb|genml.mln| and \verb|genml.acf|. This API returns an \verb|ndarray| list.

\section{Illustrative examples}

\subsection{Examples of generated M-L noise}

As shown in Fig. \ref{fig:fig2}, representative examples of M-L noise sequences generated by GenML are displayed. The input parameters for these examples are $N=1$, $T=500$, $C=1$, $\tau=10$, with $\lambda$ values of 0.6 in (a), 1.2 in (b), and 1.8 in (c), respectively. Corresponding code to get the sequence data is listed below:
\begin{verbatim}
    import genml
    xi1 = genml.mln(1, 500, 1, 0.6, 10)
    xi2 = genml.mln(1, 500, 1, 1.2, 10)
    xi3 = genml.mln(1, 500, 1, 1.8, 10)
\end{verbatim}
Following this, to quantitatively demonstrate the accuracy of M-L noise generated by GenML, we will validate the software's effectiveness from two perspectives: analysis of the autocorrelation function and examination of diffusion behavior driven by M-L noise.

\subsection{Autocorrelation function analysis}

Utilizing the two APIs provided by GenML, i.e., \verb|genml.acf| and \verb|genml.acft|, we can readily obtain both the calculated and theoretical values of the M-L noise autocorrelation function under various parameters. Fig. \ref{fig:fig3} presents an illustrative comparison between the calculated and theoretical values of autocorrelation functions for different parameter sets, where the calculated values of each set are obtained by an average of 1000 independent noise sequences. It can be observed that the calculated values (dots) align closely with the theoretical ones (lines), which directly attests to our software's capability to effectively generate M-L noise across a range of input parameters.

Since GenML is based on Fourier filtering for generating M-L noise sequences, it is important to recognize its inherent limitations. As introduced in previous foundational studies \cite{Makse, PengCK}, the effective correlation range achievable by Fourier filtering methods is inherently constrained by the finite size of generated sequences. This constraint limits the method's ability to maintain correlations over extended sequence lengths, especially as the values of the long-time autocorrelation function approach zero.

Our empirical investigations reveal that when the theoretical Mittag-Leffler autocorrelation function values approach zero in the long-time region, the calculated function values struggle to precisely converge to the theoretical values, instead exhibiting slight oscillations around zero. This nuanced behavior highlights the importance of understanding the limitations of Fourier filtering in applications where precise long-range correlations are crucial. Users of the GenML software should be particularly cautious in scenarios where the long-time autocorrelation values might approach zero.

\subsection{Examination of diffusion behavior driven by M-L noise}

\begin{figure}[t]
\centering
\includegraphics[width=8.2cm]{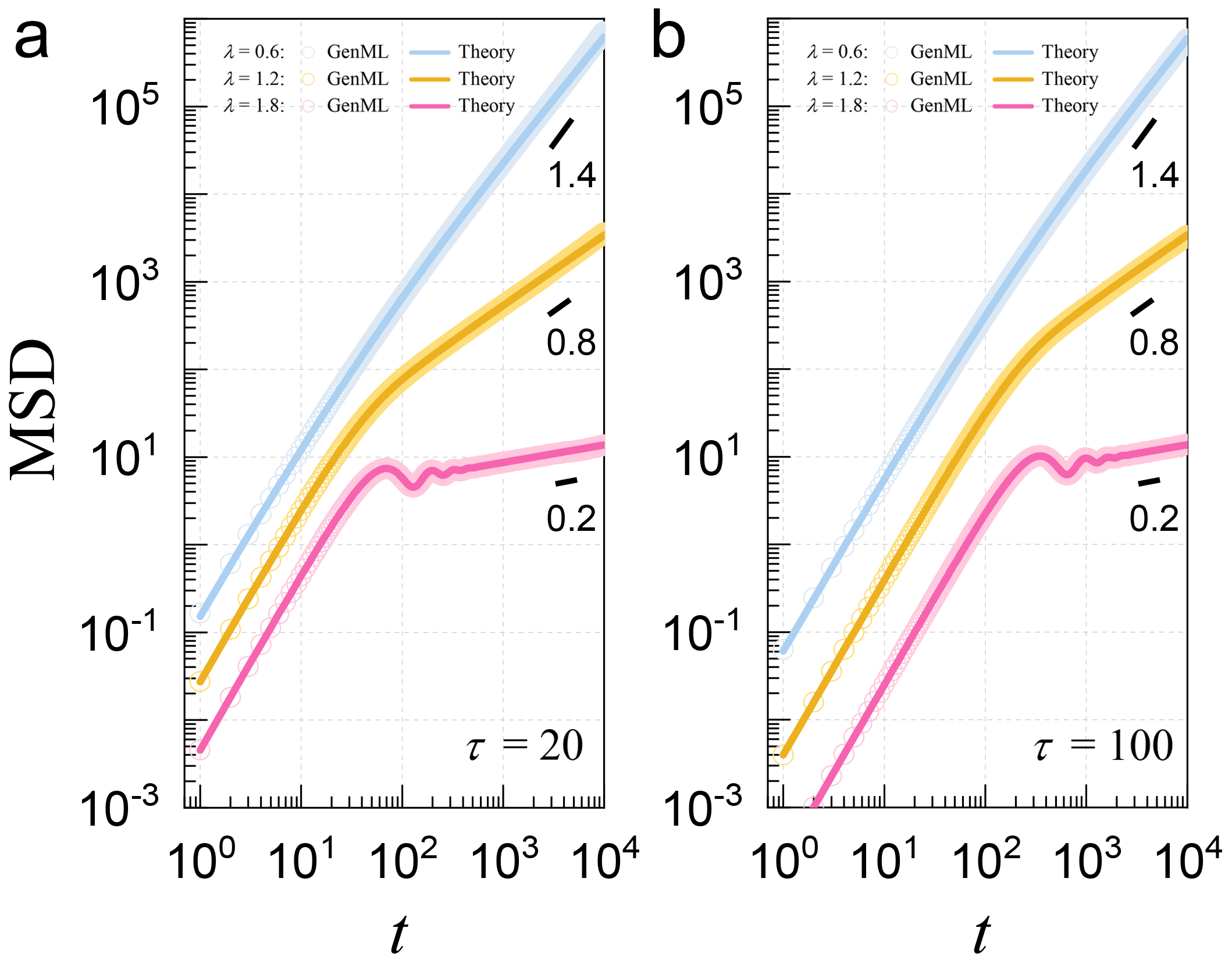}
\caption{Comparison of calculated (dots) and theoretical (lines) MSDs for the diffusion behavior driven by M-L noise. (a) and (b) show the MSDs in the log-log scale for $\lambda$ values of 0.6, 1.2, and 1.8 at $\tau$ values of 20 and 100, respectively. }
\label{fig:fig4}
\end{figure}

The anomalous diffusion of a random walker driven by M-L noise has been theoretically investigated by Vi\~{n}ales et al \cite{Vinales,Vinales2014}. Building on this theoretical foundation, we can validate the efficacy of GenML by quantitatively analyzing the diffusion behavior driven by generated M-L noise. For that purpose, the Langevin equation is utilized to obtain the trajectory $\{ x(t)\}$ of random walker driven by M-L noise $\{ \xi(t)\}$, which is expressed as:
\begin{equation}
\frac{{\rm d}x(t)}{{\rm d}t} = \xi (t).
\end{equation}
After that, the mean squared displacement (MSD) of $\{ x(t)\}$ can be calculated as:
\begin{equation}
{\rm MSD}(t) = \langle [x(t) - x(0)]^2\rangle.
\end{equation}
On the other hand, corresponding theoretical value of MSD can be given by \cite{Allen}:
\begin{equation}\label{eq:eq6}
{\rm MSD}(t)=2 \int_0^t(t-s) C(s) {\rm d} s.
\end{equation}
By substituting Eq. (\ref{eq:eq1}) into Eq. (\ref{eq:eq6}) and performing numerical integration, the theoretical value of MSD can be obtained. As illustrated in Fig. \ref{fig:fig4}, the calculated (dots) and theoretical (lines) MSDs are displayed for $\lambda$ values of 0.6, 1.2, and 1.8, at $\tau$ values of 20 (a) and 100 (b), respectively. It is evident that the calculated values closely align with the theoretical ones across different parameter sets. Moreover, as indicated by the black short lines, the slope of the long-time MSD in the log-log scale is consistent with the theoretical value $2-\lambda$ provided in Ref. \cite{Vinales2014}, collectively affirming the efficacy of our software in generating M-L noise.

\section{Impact}

To the best of our knowledge, GenML is the first tool capable of directly generating Mittag-Leffler correlated noise. This advancement is not merely technical but opens up new pathways for research and application that were previously constrained by the lack of practical generation methods for M-L noise. Beyond theoretical modeling, GenML significantly enhances the ability to incorporate M-L noise into numerical simulations and data-driven methods. This expansion means that complex systems related to M-L noise can now be described and analyzed more accurately in simulations, reflecting real-world conditions more closely than ever before. Additionally, the generation of extensive noise datasets enables the development of effective machine learning models for describing complex systems. By bridging the gap between theoretical noise models and practical application, GenML facilitates a more comprehensive understanding of complex systems across various fields, including physics, biology, and finance.

\section{Conclusions}

In this work, we have introduced GenML, a Python library designed for the generation of Mittag-Leffler correlated noise. This software addresses a critical gap in the field by providing a direct method to generate this type of noise, which is crucial for its versatility in modeling a wide range of phenomena in complex systems. The effectiveness of GenML is validated through detailed analyses of autocorrelation functions and diffusion behaviors. This validation process underscores our commitment to ensuring that GenML serves as a reliable tool for researchers and developers. We anticipate that GenML will set the stage for future discoveries and innovations, thereby contributing to a more comprehensive understanding of complex systems.\\

\section*{Acknowledgements}
We thank Wentao Ju for helpful discussions This work is supported by the National Natural Science Foundation of China (Grant No. 12104147) and the Fundamental Research Funds for the Central Universities.

\bibliography{Bibliography}

\providecommand{\noopsort}[1]{}\providecommand{\singleletter}[1]{#1}%
\begin{thebibliography}{36}%
\makeatletter
\providecommand \@ifxundefined [1]{%
 \@ifx{#1\undefined}
}%
\providecommand \@ifnum [1]{%
 \ifnum #1\expandafter \@firstoftwo
 \else \expandafter \@secondoftwo
 \fi
}%
\providecommand \@ifx [1]{%
 \ifx #1\expandafter \@firstoftwo
 \else \expandafter \@secondoftwo
 \fi
}%
\providecommand \natexlab [1]{#1}%
\providecommand \enquote  [1]{``#1''}%
\providecommand \bibnamefont  [1]{#1}%
\providecommand \bibfnamefont [1]{#1}%
\providecommand \citenamefont [1]{#1}%
\providecommand \href@noop [0]{\@secondoftwo}%
\providecommand \href [0]{\begingroup \@sanitize@url \@href}%
\providecommand \@href[1]{\@@startlink{#1}\@@href}%
\providecommand \@@href[1]{\endgroup#1\@@endlink}%
\providecommand \@sanitize@url [0]{\catcode `\\12\catcode `\$12\catcode
  `\&12\catcode `\#12\catcode `\^12\catcode `\_12\catcode `\%12\relax}%
\providecommand \@@startlink[1]{}%
\providecommand \@@endlink[0]{}%
\providecommand \url  [0]{\begingroup\@sanitize@url \@url }%
\providecommand \@url [1]{\endgroup\@href {#1}{\urlprefix }}%
\providecommand \urlprefix  [0]{URL }%
\providecommand \Eprint [0]{\href }%
\providecommand \doibase [0]{https://doi.org/}%
\providecommand \selectlanguage [0]{\@gobble}%
\providecommand \bibinfo  [0]{\@secondoftwo}%
\providecommand \bibfield  [0]{\@secondoftwo}%
\providecommand \translation [1]{[#1]}%
\providecommand \BibitemOpen [0]{}%
\providecommand \bibitemStop [0]{}%
\providecommand \bibitemNoStop [0]{.\EOS\space}%
\providecommand \EOS [0]{\spacefactor3000\relax}%
\providecommand \BibitemShut  [1]{\csname bibitem#1\endcsname}%
\let\auto@bib@innerbib\@empty
\bibitem [{\citenamefont {Franosch}\ \emph {et~al.}(2011)\citenamefont
  {Franosch}, \citenamefont {Grimm}, \citenamefont {Belushkin}, \citenamefont
  {Mor}, \citenamefont {Foffic}, \citenamefont {Forr\'{o}},\ and\ \citenamefont
  {Jeney}}]{Franosch}%
  \BibitemOpen
  \bibfield  {author} {\bibinfo {author} {\bibfnamefont {T.}~\bibnamefont
  {Franosch}}, \bibinfo {author} {\bibfnamefont {M.}~\bibnamefont {Grimm}},
  \bibinfo {author} {\bibfnamefont {M.}~\bibnamefont {Belushkin}}, \bibinfo
  {author} {\bibfnamefont {F.~M.}\ \bibnamefont {Mor}}, \bibinfo {author}
  {\bibfnamefont {G.}~\bibnamefont {Foffic}}, \bibinfo {author} {\bibfnamefont
  {L.}~\bibnamefont {Forr\'{o}}},\ and\ \bibinfo {author} {\bibfnamefont
  {S.}~\bibnamefont {Jeney}},\ }\bibfield  {title} {\bibinfo {title}
  {Resonances arising from hydrodynamic memory in {B}rownian motion},\
  }\href@noop {} {\bibfield  {journal} {\bibinfo  {journal} {Nature}\ }\textbf
  {\bibinfo {volume} {478}},\ \bibinfo {pages} {85} (\bibinfo {year}
  {2011})}\BibitemShut {NoStop}%
\bibitem [{\citenamefont {Jhawar}\ \emph {et~al.}(2020)\citenamefont {Jhawar},
  \citenamefont {Morris}, \citenamefont {Amith-Kumar}, \citenamefont {Raj},
  \citenamefont {Rogers}, \citenamefont {Rajendran},\ and\ \citenamefont
  {Guttal}}]{Jhawar}%
  \BibitemOpen
  \bibfield  {author} {\bibinfo {author} {\bibfnamefont {J.}~\bibnamefont
  {Jhawar}}, \bibinfo {author} {\bibfnamefont {R.~G.}\ \bibnamefont {Morris}},
  \bibinfo {author} {\bibfnamefont {U.~R.}\ \bibnamefont {Amith-Kumar}},
  \bibinfo {author} {\bibfnamefont {M.~D.}\ \bibnamefont {Raj}}, \bibinfo
  {author} {\bibfnamefont {T.}~\bibnamefont {Rogers}}, \bibinfo {author}
  {\bibfnamefont {H.}~\bibnamefont {Rajendran}},\ and\ \bibinfo {author}
  {\bibfnamefont {V.}~\bibnamefont {Guttal}},\ }\bibfield  {title} {\bibinfo
  {title} {Noise-induced schooling of fish},\ }\href@noop {} {\bibfield
  {journal} {\bibinfo  {journal} {Nat. Phys.}\ }\textbf {\bibinfo {volume}
  {16}},\ \bibinfo {pages} {488} (\bibinfo {year} {2020})}\BibitemShut
  {NoStop}%
\bibitem [{\citenamefont {Plenio}\ and\ \citenamefont {Huelga}(2002)}]{Plenio}%
  \BibitemOpen
  \bibfield  {author} {\bibinfo {author} {\bibfnamefont {M.~B.}\ \bibnamefont
  {Plenio}}\ and\ \bibinfo {author} {\bibfnamefont {S.~F.}\ \bibnamefont
  {Huelga}},\ }\bibfield  {title} {\bibinfo {title} {Entangled light from white
  noise},\ }\href@noop {} {\bibfield  {journal} {\bibinfo  {journal} {Phys.
  Rev. Lett.}\ }\textbf {\bibinfo {volume} {88}},\ \bibinfo {pages} {197901}
  (\bibinfo {year} {2002})}\BibitemShut {NoStop}%
\bibitem [{\citenamefont {Ghanta}\ \emph {et~al.}(2017)\citenamefont {Ghanta},
  \citenamefont {Neu},\ and\ \citenamefont {Teitsworth}}]{Ghanta}%
  \BibitemOpen
  \bibfield  {author} {\bibinfo {author} {\bibfnamefont {A.}~\bibnamefont
  {Ghanta}}, \bibinfo {author} {\bibfnamefont {J.~C.}\ \bibnamefont {Neu}},\
  and\ \bibinfo {author} {\bibfnamefont {S.}~\bibnamefont {Teitsworth}},\
  }\bibfield  {title} {\bibinfo {title} {Fluctuation loops in noise-driven
  linear dynamical systems},\ }\href@noop {} {\bibfield  {journal} {\bibinfo
  {journal} {Phys. Rev. E}\ }\textbf {\bibinfo {volume} {95}},\ \bibinfo
  {pages} {032128} (\bibinfo {year} {2017})}\BibitemShut {NoStop}%
\bibitem [{\citenamefont {Bernido}\ and\ \citenamefont
  {Carpio-Bernido}(2012)}]{BERNIDO}%
  \BibitemOpen
  \bibfield  {author} {\bibinfo {author} {\bibfnamefont {C.~C.}\ \bibnamefont
  {Bernido}}\ and\ \bibinfo {author} {\bibfnamefont {M.~V.}\ \bibnamefont
  {Carpio-Bernido}},\ }\bibfield  {title} {\bibinfo {title} {White noise
  analysis: Some applications in complex systems, biophysics and quantum
  mechanics},\ }\href@noop {} {\bibfield  {journal} {\bibinfo  {journal} {Int.
  J. Mod. Phys. B}\ }\textbf {\bibinfo {volume} {26}},\ \bibinfo {pages}
  {1230014} (\bibinfo {year} {2012})}\BibitemShut {NoStop}%
\bibitem [{\citenamefont {Chun}\ \emph {et~al.}(2018)\citenamefont {Chun},
  \citenamefont {Durang},\ and\ \citenamefont {Noh}}]{Chun}%
  \BibitemOpen
  \bibfield  {author} {\bibinfo {author} {\bibfnamefont {H.-M.}\ \bibnamefont
  {Chun}}, \bibinfo {author} {\bibfnamefont {X.}~\bibnamefont {Durang}},\ and\
  \bibinfo {author} {\bibfnamefont {J.~D.}\ \bibnamefont {Noh}},\ }\bibfield
  {title} {\bibinfo {title} {Emergence of nonwhite noise in {L}angevin dynamics
  with magnetic {L}orentz force},\ }\href@noop {} {\bibfield  {journal}
  {\bibinfo  {journal} {Phys. Rev. E}\ }\textbf {\bibinfo {volume} {97}},\
  \bibinfo {pages} {032117} (\bibinfo {year} {2018})}\BibitemShut {NoStop}%
\bibitem [{\citenamefont {Zhou}(2009)}]{Zhou}%
  \BibitemOpen
  \bibfield  {author} {\bibinfo {author} {\bibfnamefont {X.}~\bibnamefont
  {Zhou}},\ }\bibfield  {title} {\bibinfo {title} {Cooperative atomic
  scattering of light from a laser with a colored noise spectrum},\ }\href@noop
  {} {\bibfield  {journal} {\bibinfo  {journal} {Phys. Rev. A}\ }\textbf
  {\bibinfo {volume} {80}},\ \bibinfo {pages} {023818} (\bibinfo {year}
  {2009})}\BibitemShut {NoStop}%
\bibitem [{\citenamefont {Kamenev}\ \emph {et~al.}(2008)\citenamefont
  {Kamenev}, \citenamefont {Meerson},\ and\ \citenamefont
  {Shklovskii}}]{Kamenev}%
  \BibitemOpen
  \bibfield  {author} {\bibinfo {author} {\bibfnamefont {A.}~\bibnamefont
  {Kamenev}}, \bibinfo {author} {\bibfnamefont {B.}~\bibnamefont {Meerson}},\
  and\ \bibinfo {author} {\bibfnamefont {B.}~\bibnamefont {Shklovskii}},\
  }\bibfield  {title} {\bibinfo {title} {How colored environmental noise
  affects population extinction},\ }\href@noop {} {\bibfield  {journal}
  {\bibinfo  {journal} {Phys. Rev. Lett.}\ }\textbf {\bibinfo {volume} {101}},\
  \bibinfo {pages} {268103} (\bibinfo {year} {2008})}\BibitemShut {NoStop}%
\bibitem [{\citenamefont {Kazakevi\v{c}ius}\ and\ \citenamefont
  {Ruseckas}(2015)}]{Kazakevicius}%
  \BibitemOpen
  \bibfield  {author} {\bibinfo {author} {\bibfnamefont {R.}~\bibnamefont
  {Kazakevi\v{c}ius}}\ and\ \bibinfo {author} {\bibfnamefont {J.}~\bibnamefont
  {Ruseckas}},\ }\bibfield  {title} {\bibinfo {title} {Power law statistics in
  the velocity fluctuations of {B}rownian particle in inhomogeneous media and
  driven by colored noise},\ }\href@noop {} {\bibfield  {journal} {\bibinfo
  {journal} {J. Stat. Mech. Theor. Exp.}\ }\textbf {\bibinfo {volume} {2015}},\
  \bibinfo {pages} {P02021} (\bibinfo {year} {2015})}\BibitemShut {NoStop}%
\bibitem [{\citenamefont {Vi{\~{n}}ales}\ and\ \citenamefont
  {Desp\'{o}sito}(2007)}]{Vinales}%
  \BibitemOpen
  \bibfield  {author} {\bibinfo {author} {\bibfnamefont {A.~D.}\ \bibnamefont
  {Vi{\~{n}}ales}}\ and\ \bibinfo {author} {\bibfnamefont {M.~A.}\ \bibnamefont
  {Desp\'{o}sito}},\ }\bibfield  {title} {\bibinfo {title} {Anomalous diffusion
  induced by a {M}ittag-{L}effler correlated noise},\ }\href@noop {} {\bibfield
   {journal} {\bibinfo  {journal} {Phys. Rev. E}\ }\textbf {\bibinfo {volume}
  {75}},\ \bibinfo {pages} {042102} (\bibinfo {year} {2007})}\BibitemShut
  {NoStop}%
\bibitem [{\citenamefont {Vi{\~{n}}ales}\ and\ \citenamefont
  {Paissan}(2014)}]{Vinales2014}%
  \BibitemOpen
  \bibfield  {author} {\bibinfo {author} {\bibfnamefont {A.~D.}\ \bibnamefont
  {Vi{\~{n}}ales}}\ and\ \bibinfo {author} {\bibfnamefont {G.~H.}\ \bibnamefont
  {Paissan}},\ }\bibfield  {title} {\bibinfo {title} {Velocity autocorrelation
  of a free particle driven by a {M}ittag-{L}effler noise: {F}ractional
  dynamics and temporal behaviors},\ }\href@noop {} {\bibfield  {journal}
  {\bibinfo  {journal} {Phys. Rev. E}\ }\textbf {\bibinfo {volume} {90}},\
  \bibinfo {pages} {062103} (\bibinfo {year} {2014})}\BibitemShut {NoStop}%
\bibitem [{\citenamefont {Fa}(2020)}]{Fa}%
  \BibitemOpen
  \bibfield  {author} {\bibinfo {author} {\bibfnamefont {K.~S.}\ \bibnamefont
  {Fa}},\ }\bibfield  {title} {\bibinfo {title} {Fractional oscillator noise
  and its applications},\ }\href@noop {} {\bibfield  {journal} {\bibinfo
  {journal} {Int. J. Mod. Phys. B}\ }\textbf {\bibinfo {volume} {34}},\
  \bibinfo {pages} {2050234} (\bibinfo {year} {2020})}\BibitemShut {NoStop}%
\bibitem [{\citenamefont {Laas}\ and\ \citenamefont {Mankin}(2015)}]{Laas}%
  \BibitemOpen
  \bibfield  {author} {\bibinfo {author} {\bibfnamefont {K.}~\bibnamefont
  {Laas}}\ and\ \bibinfo {author} {\bibfnamefont {R.}~\bibnamefont {Mankin}},\
  }\bibfield  {title} {\bibinfo {title} {Resonance of {B}rownian vortices in
  viscoelastic shear flows},\ }in\ \href@noop {} {\emph {\bibinfo {booktitle}
  {AIP Conf. Proc.}}},\ Vol.\ \bibinfo {volume} {1684}\ (\bibinfo
  {organization} {AIP Publishing},\ \bibinfo {year} {2015})\BibitemShut
  {NoStop}%
\bibitem [{\citenamefont {Cairano}\ \emph {et~al.}(2021)\citenamefont
  {Cairano}, \citenamefont {Stamm},\ and\ \citenamefont
  {Calandrini}}]{Cairano}%
  \BibitemOpen
  \bibfield  {author} {\bibinfo {author} {\bibfnamefont {L.~D.}\ \bibnamefont
  {Cairano}}, \bibinfo {author} {\bibfnamefont {B.}~\bibnamefont {Stamm}},\
  and\ \bibinfo {author} {\bibfnamefont {V.}~\bibnamefont {Calandrini}},\
  }\bibfield  {title} {\bibinfo {title} {Subdiffusive-{B}rownian crossover in
  membrane proteins: {A} generalized {L}angevin equation-based approach},\
  }\href@noop {} {\bibfield  {journal} {\bibinfo  {journal} {Biophys. J.}\
  }\textbf {\bibinfo {volume} {120}},\ \bibinfo {pages} {4722} (\bibinfo {year}
  {2021})}\BibitemShut {NoStop}%
\bibitem [{\citenamefont {Umamaheswaria}\ \emph {et~al.}(2020)\citenamefont
  {Umamaheswaria}, \citenamefont {Balachandrana},\ and\ \citenamefont
  {Annapoorania}}]{Umamaheswaria}%
  \BibitemOpen
  \bibfield  {author} {\bibinfo {author} {\bibfnamefont {P.}~\bibnamefont
  {Umamaheswaria}}, \bibinfo {author} {\bibfnamefont {K.}~\bibnamefont
  {Balachandrana}},\ and\ \bibinfo {author} {\bibfnamefont {N.}~\bibnamefont
  {Annapoorania}},\ }\bibfield  {title} {\bibinfo {title} {Existence and
  stability results for {C}aputo fractional stochastic differential equations
  with {L}\'{e}vy noise},\ }\href@noop {} {\bibfield  {journal} {\bibinfo
  {journal} {Filomat}\ }\textbf {\bibinfo {volume} {34}},\ \bibinfo {pages}
  {1739} (\bibinfo {year} {2020})}\BibitemShut {NoStop}%
\bibitem [{\citenamefont {Milotti}(2005)}]{Milotti}%
  \BibitemOpen
  \bibfield  {author} {\bibinfo {author} {\bibfnamefont {E.}~\bibnamefont
  {Milotti}},\ }\bibfield  {title} {\bibinfo {title} {Exact numerical
  simulation of power-law noises},\ }\href@noop {} {\bibfield  {journal}
  {\bibinfo  {journal} {Phys. Rev. E}\ }\textbf {\bibinfo {volume} {72}},\
  \bibinfo {pages} {056701} (\bibinfo {year} {2005})}\BibitemShut {NoStop}%
\bibitem [{\citenamefont {Kasdin}(1995)}]{Kasdin}%
  \BibitemOpen
  \bibfield  {author} {\bibinfo {author} {\bibfnamefont {N.~J.}\ \bibnamefont
  {Kasdin}},\ }\bibfield  {title} {\bibinfo {title} {Discrete simulation of
  colored noise and stochastic processes and $1/f^\alpha$ power law noise
  generation},\ }\href@noop {} {\bibfield  {journal} {\bibinfo  {journal}
  {Proc. IEEE}\ }\textbf {\bibinfo {volume} {83}},\ \bibinfo {pages} {802}
  (\bibinfo {year} {1995})}\BibitemShut {NoStop}%
\bibitem [{\citenamefont {Bykhovsky}(2018)}]{Bykhovsky}%
  \BibitemOpen
  \bibfield  {author} {\bibinfo {author} {\bibfnamefont {D.}~\bibnamefont
  {Bykhovsky}},\ }\bibfield  {title} {\bibinfo {title} {Mathematica code for
  numerical generation of random process with given distribution and
  exponential autocorrelation function},\ }\href@noop {} {\bibfield  {journal}
  {\bibinfo  {journal} {SoftwareX}\ }\textbf {\bibinfo {volume} {8}},\ \bibinfo
  {pages} {18} (\bibinfo {year} {2018})}\BibitemShut {NoStop}%
\bibitem [{\citenamefont {Stella}\ \emph {et~al.}(2014)\citenamefont {Stella},
  \citenamefont {Lorenz},\ and\ \citenamefont {Kantorovich}}]{Stella}%
  \BibitemOpen
  \bibfield  {author} {\bibinfo {author} {\bibfnamefont {L.}~\bibnamefont
  {Stella}}, \bibinfo {author} {\bibfnamefont {C.~D.}\ \bibnamefont {Lorenz}},\
  and\ \bibinfo {author} {\bibfnamefont {L.}~\bibnamefont {Kantorovich}},\
  }\bibfield  {title} {\bibinfo {title} {Generalized {L}angevin equation: {A}n
  efficient approach to nonequilibrium molecular dynamics of open systems},\
  }\href@noop {} {\bibfield  {journal} {\bibinfo  {journal} {Phys. Rev. B}\
  }\textbf {\bibinfo {volume} {89}},\ \bibinfo {pages} {134303} (\bibinfo
  {year} {2014})}\BibitemShut {NoStop}%
\bibitem [{\citenamefont {Burov}\ and\ \citenamefont {Barkai}(2008)}]{Burov}%
  \BibitemOpen
  \bibfield  {author} {\bibinfo {author} {\bibfnamefont {S.}~\bibnamefont
  {Burov}}\ and\ \bibinfo {author} {\bibfnamefont {E.}~\bibnamefont {Barkai}},\
  }\bibfield  {title} {\bibinfo {title} {Critical exponent of the fractional
  langevin equation},\ }\href@noop {} {\bibfield  {journal} {\bibinfo
  {journal} {Phys. Rev. Lett.}\ }\textbf {\bibinfo {volume} {100}},\ \bibinfo
  {pages} {070601} (\bibinfo {year} {2008})}\BibitemShut {NoStop}%
\bibitem [{\citenamefont {Hollingsworth}\ and\ \citenamefont
  {Dror}(2018)}]{Hollingsworth}%
  \BibitemOpen
  \bibfield  {author} {\bibinfo {author} {\bibfnamefont {S.~A.}\ \bibnamefont
  {Hollingsworth}}\ and\ \bibinfo {author} {\bibfnamefont {R.~O.}\ \bibnamefont
  {Dror}},\ }\bibfield  {title} {\bibinfo {title} {Molecular dynamics
  simulation for all},\ }\href@noop {} {\bibfield  {journal} {\bibinfo
  {journal} {Neuron}\ }\textbf {\bibinfo {volume} {99}},\ \bibinfo {pages}
  {1129} (\bibinfo {year} {2018})}\BibitemShut {NoStop}%
\bibitem [{\citenamefont {{Mu{\~{n}}oz-Gil et al.}}(2021)}]{Munoz-Gil}%
  \BibitemOpen
  \bibfield  {author} {\bibinfo {author} {\bibfnamefont {G.}~\bibnamefont
  {{Mu{\~{n}}oz-Gil et al.}}},\ }\bibfield  {title} {\bibinfo {title}
  {Objective comparison of methods to decode anomalous diffusion},\ }\href@noop
  {} {\bibfield  {journal} {\bibinfo  {journal} {Nat. Commun.}\ }\textbf
  {\bibinfo {volume} {12}},\ \bibinfo {pages} {6253} (\bibinfo {year}
  {2021})}\BibitemShut {NoStop}%
\bibitem [{\citenamefont {Bo}\ \emph {et~al.}(2019)\citenamefont {Bo},
  \citenamefont {Schmidt}, \citenamefont {Eichhorn},\ and\ \citenamefont
  {Volpe}}]{Bo}%
  \BibitemOpen
  \bibfield  {author} {\bibinfo {author} {\bibfnamefont {S.}~\bibnamefont
  {Bo}}, \bibinfo {author} {\bibfnamefont {F.}~\bibnamefont {Schmidt}},
  \bibinfo {author} {\bibfnamefont {R.}~\bibnamefont {Eichhorn}},\ and\
  \bibinfo {author} {\bibfnamefont {G.}~\bibnamefont {Volpe}},\ }\bibfield
  {title} {\bibinfo {title} {Measurement of anomalous diffusion using recurrent
  neural networks},\ }\href@noop {} {\bibfield  {journal} {\bibinfo  {journal}
  {Phys. Rev. E}\ }\textbf {\bibinfo {volume} {100}},\ \bibinfo {pages}
  {010102} (\bibinfo {year} {2019})}\BibitemShut {NoStop}%
\bibitem [{\citenamefont {Qu}\ \emph {et~al.}(2024)\citenamefont {Qu},
  \citenamefont {Hu}, \citenamefont {Cai}, \citenamefont {Xu}, \citenamefont
  {Ke}, \citenamefont {Zhu},\ and\ \citenamefont {Huang}}]{Qu}%
  \BibitemOpen
  \bibfield  {author} {\bibinfo {author} {\bibfnamefont {X.}~\bibnamefont
  {Qu}}, \bibinfo {author} {\bibfnamefont {Y.}~\bibnamefont {Hu}}, \bibinfo
  {author} {\bibfnamefont {W.}~\bibnamefont {Cai}}, \bibinfo {author}
  {\bibfnamefont {Y.}~\bibnamefont {Xu}}, \bibinfo {author} {\bibfnamefont
  {H.}~\bibnamefont {Ke}}, \bibinfo {author} {\bibfnamefont {G.}~\bibnamefont
  {Zhu}},\ and\ \bibinfo {author} {\bibfnamefont {Z.}~\bibnamefont {Huang}},\
  }\bibfield  {title} {\bibinfo {title} {Semantic segmentation of anomalous
  diffusion using deep convolutional networks},\ }\href@noop {} {\bibfield
  {journal} {\bibinfo  {journal} {Phys. Rev. Research}\ }\textbf {\bibinfo
  {volume} {6}},\ \bibinfo {pages} {013054} (\bibinfo {year}
  {2024})}\BibitemShut {NoStop}%
\bibitem [{\citenamefont {Li}\ \emph {et~al.}(2021)\citenamefont {Li},
  \citenamefont {Yao},\ and\ \citenamefont {Huang}}]{Li}%
  \BibitemOpen
  \bibfield  {author} {\bibinfo {author} {\bibfnamefont {D.}~\bibnamefont
  {Li}}, \bibinfo {author} {\bibfnamefont {Q.}~\bibnamefont {Yao}},\ and\
  \bibinfo {author} {\bibfnamefont {Z.}~\bibnamefont {Huang}},\ }\bibfield
  {title} {\bibinfo {title} {Wave{N}et-based deep neural networks for the
  characterization of anomalous diffusion ({WADNet})},\ }\href@noop {}
  {\bibfield  {journal} {\bibinfo  {journal} {J. Phys. A Math. Theor.}\
  }\textbf {\bibinfo {volume} {54}},\ \bibinfo {pages} {404003} (\bibinfo
  {year} {2021})}\BibitemShut {NoStop}%
\bibitem [{\citenamefont {Feng}\ \emph {et~al.}(2023)\citenamefont {Feng},
  \citenamefont {Wang}, \citenamefont {Liu}, \citenamefont {Li},\ and\
  \citenamefont {Xu}}]{Feng}%
  \BibitemOpen
  \bibfield  {author} {\bibinfo {author} {\bibfnamefont {J.}~\bibnamefont
  {Feng}}, \bibinfo {author} {\bibfnamefont {X.}~\bibnamefont {Wang}}, \bibinfo
  {author} {\bibfnamefont {Q.}~\bibnamefont {Liu}}, \bibinfo {author}
  {\bibfnamefont {Y.}~\bibnamefont {Li}},\ and\ \bibinfo {author}
  {\bibfnamefont {Y.}~\bibnamefont {Xu}},\ }\bibfield  {title} {\bibinfo
  {title} {Deep learning-based parameter estimation of stochastic differential
  equations driven by fractional {B}rownian motions with measurement noise},\
  }\href@noop {} {\bibfield  {journal} {\bibinfo  {journal} {Commun. Nonlinear
  Sci. Numer. Simul.}\ }\textbf {\bibinfo {volume} {127}},\ \bibinfo {pages}
  {107589} (\bibinfo {year} {2023})}\BibitemShut {NoStop}%
\bibitem [{gen()}]{genml}%
  \BibitemOpen
  \bibfield  {title} {\bibinfo {title} {Git{H}ub, genml},\ }\href@noop {}
  {\bibinfo  {journal} {\url{https://github.com/huangzih/genml}, 2024}\
  }\BibitemShut {NoStop}%
\bibitem [{\citenamefont {Garrappa}(2015)}]{Garrappa}%
  \BibitemOpen
\bibfield  {journal} {  }\bibfield  {author} {\bibinfo {author} {\bibfnamefont
  {R.}~\bibnamefont {Garrappa}},\ }\bibfield  {title} {\bibinfo {title}
  {Numerical evaluation of two and three parameter {M}ittag-{L}effler
  functions},\ }\href@noop {} {\bibfield  {journal} {\bibinfo  {journal} {SIAM
  J. Numer. Anal.}\ }\textbf {\bibinfo {volume} {53}},\ \bibinfo {pages} {1350}
  (\bibinfo {year} {2015})}\BibitemShut {NoStop}%
\bibitem [{mit()}]{mittag-leffler}%
  \BibitemOpen
  \bibfield  {title} {\bibinfo {title} {Git{H}ub, mittag-leffler},\ }\href@noop
  {} {\bibinfo  {journal} {\url{https://github.com/khinsen/mittag-leffler},
  2017}\ }\BibitemShut {NoStop}%
\bibitem [{\citenamefont {Rjasanow}(1994)}]{Sergej}%
  \BibitemOpen
\bibfield  {journal} {  }\bibfield  {author} {\bibinfo {author} {\bibfnamefont
  {S.}~\bibnamefont {Rjasanow}},\ }\bibfield  {title} {\bibinfo {title}
  {Effective algorithms with circulant-block matrices},\ }\href@noop {}
  {\bibfield  {journal} {\bibinfo  {journal} {Linear Algebra Appl.}\ }\textbf
  {\bibinfo {volume} {202}},\ \bibinfo {pages} {55} (\bibinfo {year}
  {1994})}\BibitemShut {NoStop}%
\bibitem [{\citenamefont {Davies}\ and\ \citenamefont {Harte}(1987)}]{DAVIES}%
  \BibitemOpen
  \bibfield  {author} {\bibinfo {author} {\bibfnamefont {R.~B.}\ \bibnamefont
  {Davies}}\ and\ \bibinfo {author} {\bibfnamefont {D.~S.}\ \bibnamefont
  {Harte}},\ }\bibfield  {title} {\bibinfo {title} {Tests for {H}urst effect},\
  }\href@noop {} {\bibfield  {journal} {\bibinfo  {journal} {Biometrika}\
  }\textbf {\bibinfo {volume} {74}},\ \bibinfo {pages} {95} (\bibinfo {year}
  {1987})}\BibitemShut {NoStop}%
\bibitem [{\citenamefont {Craigmile}(2003)}]{CRAIGMILE}%
  \BibitemOpen
  \bibfield  {author} {\bibinfo {author} {\bibfnamefont {P.~F.}\ \bibnamefont
  {Craigmile}},\ }\bibfield  {title} {\bibinfo {title} {Simulating a class of
  stationary {G}aussian processes using the {D}avies–{H}arte algorithm, with
  application to long memory processes},\ }\href@noop {} {\bibfield  {journal}
  {\bibinfo  {journal} {J. Time Ser. Anal.}\ }\textbf {\bibinfo {volume}
  {24}},\ \bibinfo {pages} {505} (\bibinfo {year} {2003})}\BibitemShut
  {NoStop}%
\bibitem [{\citenamefont {McKerns}\ \emph {et~al.}(2012)\citenamefont
  {McKerns}, \citenamefont {Strand}, \citenamefont {Sullivan}, \citenamefont
  {Fang},\ and\ \citenamefont {Aivazis}}]{McKerns}%
  \BibitemOpen
  \bibfield  {author} {\bibinfo {author} {\bibfnamefont {M.~M.}\ \bibnamefont
  {McKerns}}, \bibinfo {author} {\bibfnamefont {L.}~\bibnamefont {Strand}},
  \bibinfo {author} {\bibfnamefont {T.}~\bibnamefont {Sullivan}}, \bibinfo
  {author} {\bibfnamefont {A.}~\bibnamefont {Fang}},\ and\ \bibinfo {author}
  {\bibfnamefont {M.~A.}\ \bibnamefont {Aivazis}},\ }\bibfield  {title}
  {\bibinfo {title} {Building a framework for predictive science},\ }\href@noop
  {} {\bibfield  {journal} {\bibinfo  {journal} {arXiv preprint
  arXiv:1202.1056}\ } (\bibinfo {year} {2012})}\BibitemShut {NoStop}%
\bibitem [{\citenamefont {Makse}\ \emph {et~al.}(1996)\citenamefont {Makse},
  \citenamefont {Havlin}, \citenamefont {Schwartz},\ and\ \citenamefont
  {Stanley}}]{Makse}%
  \BibitemOpen
  \bibfield  {author} {\bibinfo {author} {\bibfnamefont {H.~A.}\ \bibnamefont
  {Makse}}, \bibinfo {author} {\bibfnamefont {S.}~\bibnamefont {Havlin}},
  \bibinfo {author} {\bibfnamefont {M.}~\bibnamefont {Schwartz}},\ and\
  \bibinfo {author} {\bibfnamefont {H.~E.}\ \bibnamefont {Stanley}},\
  }\bibfield  {title} {\bibinfo {title} {Method for generating long-range
  correlations for large systems},\ }\href@noop {} {\bibfield  {journal}
  {\bibinfo  {journal} {Phys. Rev. E}\ }\textbf {\bibinfo {volume} {53}},\
  \bibinfo {pages} {5445} (\bibinfo {year} {1996})}\BibitemShut {NoStop}%
\bibitem [{\citenamefont {Peng}\ \emph {et~al.}(1991)\citenamefont {Peng},
  \citenamefont {Havlin}, \citenamefont {Schwartz},\ and\ \citenamefont
  {Stanley}}]{PengCK}%
  \BibitemOpen
  \bibfield  {author} {\bibinfo {author} {\bibfnamefont {C.}~\bibnamefont
  {Peng}}, \bibinfo {author} {\bibfnamefont {S.}~\bibnamefont {Havlin}},
  \bibinfo {author} {\bibfnamefont {M.}~\bibnamefont {Schwartz}},\ and\
  \bibinfo {author} {\bibfnamefont {H.~E.}\ \bibnamefont {Stanley}},\
  }\bibfield  {title} {\bibinfo {title} {Directed-polymer and
  ballistic-deposition growth with correlated noise},\ }\href@noop {}
  {\bibfield  {journal} {\bibinfo  {journal} {Phys. Rev. A}\ }\textbf {\bibinfo
  {volume} {44}},\ \bibinfo {pages} {R2239} (\bibinfo {year}
  {1991})}\BibitemShut {NoStop}%
\bibitem [{\citenamefont {Allen}\ and\ \citenamefont
  {Tildesley}(2017)}]{Allen}%
  \BibitemOpen
  \bibfield  {author} {\bibinfo {author} {\bibfnamefont {M.~P.}\ \bibnamefont
  {Allen}}\ and\ \bibinfo {author} {\bibfnamefont {D.~J.}\ \bibnamefont
  {Tildesley}},\ }\href@noop {} {\emph {\bibinfo {title} {Computer simulation
  of liquids}}}\ (\bibinfo  {publisher} {Oxford University Press},\ \bibinfo
  {year} {2017})\BibitemShut {NoStop}%
\end{thebibliography}%

\end{document}